\documentclass[trackchanges]{aastex701}

\usepackage{amsmath}  
\usepackage{xcolor}
\usepackage{caption}

\begin{document}

\title{How Beaming Shapes the Demographics of Ultraluminous X-ray Sources?}

\author[orcid=0000-0001-8356-2233]{Ying-Han Mao}
\affiliation{School of Astronomy and Space Science, Nanjing University, Nanjing 210023, P. R. China}
\affiliation{Key Laboratory of Modern Astronomy and Astrophysics (Nanjing University),
Ministry of Education, Nanjing 210023, P. R. China}
\email{maoyh@smail.nju.edu.cn}

\author[orcid=0000-0002-0584-8145]{Xiang-Dong Li}
\affiliation{School of Astronomy and Space Science, Nanjing University, Nanjing 210023, P. R. China}
\affiliation{Key Laboratory of Modern Astronomy and Astrophysics (Nanjing University),
Ministry of Education, Nanjing 210023, P. R. China}
\email[show]{lixd@nju.edu.cn}

\begin{abstract}
Ultraluminous X-ray sources (ULXs) are off-nuclear compact objects with apparent luminosities above $10^{39}~\rm{erg~s^{-1}}$, often exceeding the Eddington limit for stellar-mass black holes. Beaming is a commonly invoked mechanism to explain their extreme brightness, and the dependence of the beaming factor on accretion rate is a critical parameter. In this work, we investigate how different beaming prescriptions affect the predicted properties of ULX populations. Using binary population synthesis, we construct synthetic X-ray luminosity functions (XLFs) for both classical and log-modified beaming models at solar and sub-solar metallicities. The classical model predicts a larger intrinsic number of bright ULXs, but strong beaming reduces their observable fraction, resulting in fewer visible ULXs compared to the log-modified model. The log-modified prescription yields a shallower slope at high-luminosity, aligning better with observed XLFs, and increases the fraction of observable neutron star ULXs above $10^{39}~\rm{erg~s^{-1}}$. These results underscore the significant role of the beaming law in shaping ULX statistical distributions  and assessing neutron star contributions to the population.

\end{abstract}


\section{Introduction} \label{Introduction}

Ultraluminous X-ray sources (ULXs) are point-like X-ray sources with luminosities exceeding $10^{39} ~\rm{erg~s^{-1}}$, with the assumption of isotropic emission. This exceeds the classical Eddington luminosity for $\sim 10~{\rm M_{\odot}}$ black holes (BHs) and challenges standard accretion models \citep{Fabbiano1989,Kaaret2017}. Early interpretations suggested that ULXs host accreting intermediate-mass black holes (IMBHs) with masses in the range of $10^2$–$10^4~{\rm M_\odot}$ \citep{Colbert1999,Miller2003}. However, growing observational evidence, particularly the detection of pulsating ULXs \citep{Bachetti2014}, indicates that many ULXs may instead be powered by stellar-mass compact objects. Alternative models include relativistic beaming from microquasar-like jets \citep{Kording2002}, magnetic field-enhanced super-Eddington accretion onto neutron stars (NSs) \citep[e.g.,][]{Mushtukov2015}, and anisotropic emission due to geometrical collimation \citep{King2001}.

A prominent explanation for the luminosities of ULXs is the beaming model \citep{King2001,King2009,King2017,King2019,King2024}, which assumes that radiation is preferentially emitted along narrow funnels formed by supercritical accretion disks. This scenario allows the the apparent isotropic luminosity to significantly exceed the local Eddington limit without violating radiation pressure constraints \citep{King2001,Poutanen2007}. Some general-relativistic radiative magnetohydrodynamic simulations also support the idea that radiation will experience strong beaming \citep[e.g.,][]{Abarca2021,Kayanikhoo2025}. At the same time, several observational results point to cases where strong beaming may be limited, such as the minimal geometric collimation inferred for ULX-1 \citep{Binder2018} and the large pulsed fractions observed in neutron-star ULXs, which can constrain the degree of beaming \citep{Mushtukov2021}.

Although beaming may not apply universally, it remains a viable explanation for a substantial subset of stellar-mass ULXs. The beaming factor $b<1$ quantifies the degree of beaming, defined as $b=\Omega/4\pi$, where $\Omega$ is the solid angle of the collimated emission cone. The apparent isotropic luminosity is then $L_{\rm iso} = L_{\rm int}/b$, with $L_{\rm int}$ being the intrinsic luminosity. High mass transfer rates can lead to strongly anisotropic emission, enabling stellar-mass BHs or NSs to appear as ULXs. Beaming also introduces observational selection effects, as only systems with emission cones aligned toward observers are detectable as ULXs \citep{Sutton2013}.

The beaming factor $b$ is  critical, influencing both the inferred intrinsic properties of ULXs and their detectability. The original beaming factor formulation proposed by \citet{King2009} links $b$ with the dimensionless accretion rate $\dot{m} = \dot{M}/\dot{M}_{\rm Edd}$, where $\dot{M}_{\rm Edd} = 1.3 \times 10^{18} \times (M_1/{\rm M_\odot})~\rm{g~s^{-1}}$ denotes the Eddington accretion rate for a compact object of mass $M_1$.
By combining beaming with blackbody emission in the supercritical regime \citep{Shakura1973} and comparing theoretical predictions with the observed soft-component luminosity–temperature relation \citep{Kajava2008}, \citet{King2009} derived the following expression for the beaming factor:
\begin{equation}
b = \left(\frac{8.5}{\dot{m}}\right)^2 (1 + \ln \dot{m})^2
\end{equation}
This expression was later simplified to the more commonly used form，
\begin{equation}
b = \left(\frac{8.5}{\dot{m}}\right)^2
\end{equation}
assuming that the logarithmic correction term $\sim 1$. However, during the thermal-timescale mass transfer phase, with mass transfer rates can exceed $100 \dot{M}_{\rm Edd}$, the logarithmic correction factor $l = 1 + \ln \dot{m}$ becomes significant and cannot be neglected. This rapid transfer stage is widely regarded as a likely evolutionary stage through which compact objects can be observed as ULXs \citep{King2001}. Accordingly, different  beaming factor prescriptions (Eqs. 1 and 2) imply different degrees of beaming, with smaller $b$ indicating more extreme collimation, directly influencing the predicted ULX populations.

Many earlier studies have used binary population synthesis (BPS) to explore the ULX population. Wind-fed systems have been shown to contribute substantially to the population, particularly in binaries with massive or evolved donors \citep{Wiktorowicz2021, Zuo2021}. Other work has examined Roche-lobe overflow channels, indicating that highly super-Eddington mass transfer can produce neutron-star ULXs \citep{Shao2015, Shao2019}. The relative contributions of NS and BH accretors, as well as their dependence on star-formation history, have also been investigated \citep{Wiktorowicz2017, Wiktorowicz2019, Kuranov2021}.

In this work, we examine how different beaming factor prescriptions affect the predicted properties of ULX populations. By combining binary population synthesis with different $b(\dot{m})$ formulations, we construct synthetic ULX samples and compare their luminosity distributions with observations, assessing the sensitivity of population predictions to the assumed beaming models. 
The paper is structured as follows: Section~\ref{sec:Method} describes the models and assumptions, Section~\ref{sec:Results} presents the predicted populations, and Section~\ref{sec:Conclusion} summarizes our findings.

\section{Method} \label{sec:Method}
\subsection{Beaming Model} \label{subsec:2.1}

\begin{figure}[ht!]
\centering
\includegraphics[height=18cm]{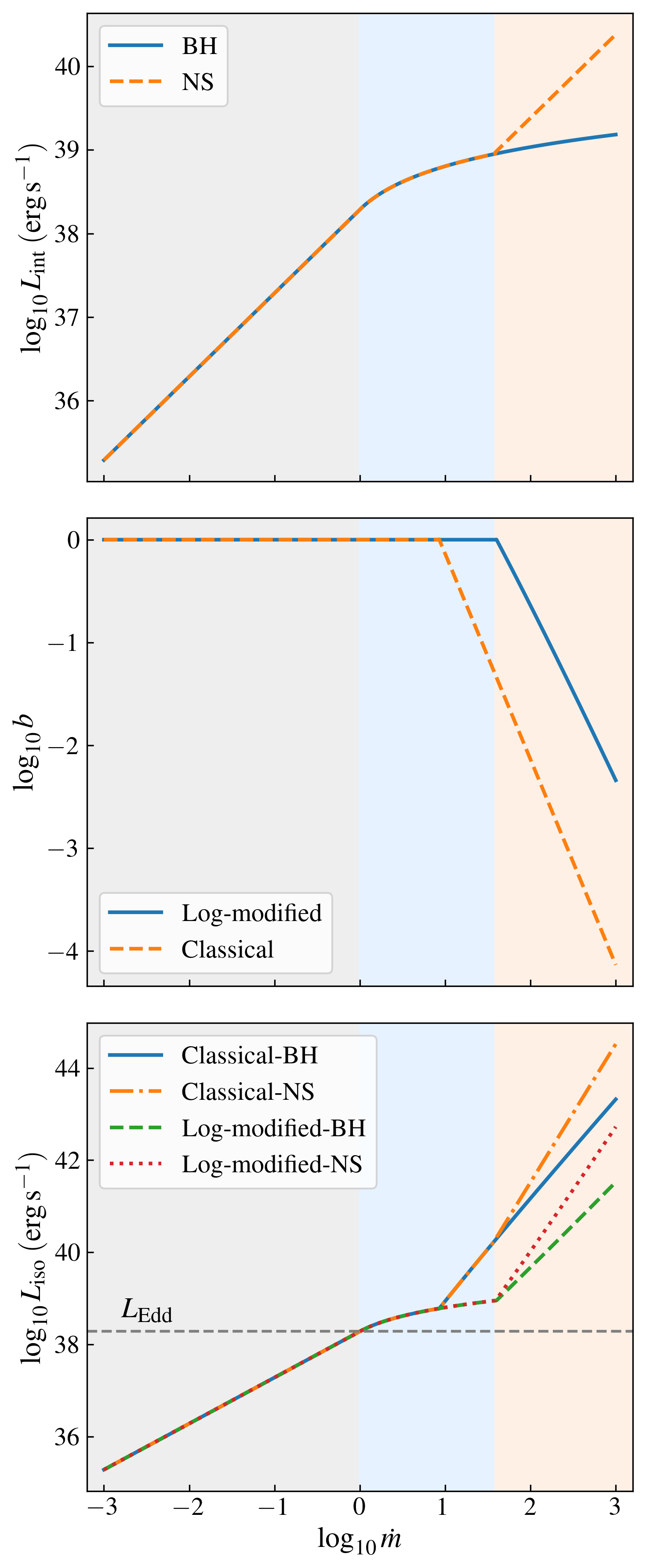}  
\caption{Comparison of the classical and log-modified beaming prescriptions as a function of the dimensionless mass accretion rate $\dot{m}$. The gray, blue, and orange shaded regions correspond to $\dot{m}<1$, $1<\dot{m}<38.5$, and $\dot{m}>38.5$, respectively.
Top panel: Intrinsic luminosity $L_{\rm int}$ as a function of $\dot{m}$. A clear transition from linear scaling at $\dot{m} \leq 1$ to logarithmic scaling at $\dot{m} > 1$ is visible. The solid and dashed lines represent BH and NS of the same mass.
Middle panel: Beaming factor $b(\dot{m})$ for the classical (dashed line) and log-modified (solid line) prescriptions. The classical model adopts a constant $x=1$ with $\dot{m}_{\rm crit}=8.5$, whereas in the log-modified model $x$ varies with $\dot{m}$, giving $\dot{m}_{\rm crit}=38.5$.
Bottom panel: Observed isotropic luminosity $L_{\rm iso} = L_{\rm int} / b$ for the two models. At low accretion rates ($\dot{m} < 8.5$), the two curves coincide. At high accretion rates, the log-modified model predicts a significantly lower $L_{\rm iso}$ than the classical model, with the divergence between the two curves increasing rapidly with $\dot{m}$, consistent with the behavior of $b(\dot{m})$. Solid and dash-dotted lines denote BH and NS in the classical model; dashed and dotted lines show the log-modified case. All curves are shown for a central compact object mass of $M_1 = 1.4~{\rm M_{\odot}}$, taken as a typical NS mass and serving as a reference case.
}
\label{fig:figure1}
\end{figure}

The supercritical accretion model proposed by \citet{Shakura1973} (hereafter referred to as the SS model) serves as a widely adopted framework for characterizing the luminosity of an accreting object as a function of its mass accretion rate $\dot{m}$. In the SS model's depiction of supercritical accretion onto BHs, the accretion disk is locally Eddington-limited across all radii, predicting the formation of strong outflows within the spherization radius. It is crucial to note that the SS model assumes outward energy transport primarily via radiation and overlooks energy advection by the inflowing material. Recent numerical simulations, however, reveal that advection can alter the spherization radius and, consequently, influence the intrinsic luminosity \citep{Abarca2021,Kayanikhoo2025}. The onset of significant advection may occur at accretion rates as low as $\sim 20 \dot{M}_{\rm Edd}$ \citep{Abarca2021} or as high as $\sim 1000 \dot{M}_{\rm Edd}$ \citep{Chashkina2019}. In this scenario, both matter and energy are carried inward via advection, with a portion of the advected luminosity ultimately being released within the inner regions of the disk. Conversely, if the accretor is a NS, the presence of a solid surface means that most of the energy is released at the NS's surface, potentially resulting in stronger emission compared to BHs. Given the differences in accretion between BHs and NSs, we formulate the intrinsic bolometric luminosity  $L_{\rm {int}}$ as follows:
\begin{equation}
L_{\text{int}} = L_{\rm Edd} \cdot \left\{
    \begin{aligned}
        & \dot{m}, \quad & \dot{m} \leq 1   \\
        & 1 + \ln \dot{m}, \quad & \dot{m} > 1, \ \rm{BH}\\
        &\max (1 + \ln \dot{m}, 0.125 \dot{m}), \quad & \dot{m} > 1, \ \rm{NS},
    \end{aligned}
\right.
\end{equation}
where $L_{\rm Edd} = 1.3 \times 10^{38}\times (M_1/{\rm M_\odot})~\rm{erg~s^{-1}} $ is the Eddington luminosity of a compact object with mass $M_1$. This formulation highlights a transition from a linear scaling of luminosity with $\dot{m}$ at sub-Eddington regime ($\dot{m} \leq 1$) to a logarithmic dependence in the super-Eddington accretion rates ($\dot{m} > 1$). In the super-Eddington regime, we further differentiate between BHs and NSs by employing distinct intrinsic luminosity formulations. For BHs, we use the traditional expression from the SS model. For NSs, we assume the intrinsic luminosity follows $L_{\rm int} \propto \max \left( 1 + \ln \dot{m}, 0.125\dot{m} \right)$, with the $0.125\dot{m}$ term  inspired by the reflection efficiency reported by \citet{Abarca2021}. This formulation starts to diverge from the BH case at $\dot{m} \simeq 38.5$; for $\dot{m} < 38.5$, the expressions for NSs and BHs are identical, whereas at higher accretion rates, NSs exhibit greater intrinsic luminosity, as illustrated in the top panel of Figure~\ref{fig:figure1}.

The intrinsic luminosity is related to the observed luminosity through the beaming factor $b$, which quantifies the degree of geometrical collimation of the emission and plays a central role in shaping the observed properties and statistical distribution of ULXs, such as the X-ray luminosity function (XLF). Following Eq.~(8) of \citet{King2009}, the beaming factor $b$ can be expressed as:
\begin{equation}
b = \left(\frac{8.5}{\dot{m}}\right)^2 x,
\end{equation}
where
\begin{equation}
x = \frac{l^2}{p \, \overline{r}^2},
\end{equation}
where $l = 1 + \ln \dot{m}$, $p$ is a factor accounting for deviations from spherical symmetry in the emission geometry, and $\overline{r} \sim 1$ is a dimensionless parameter related to the blackbody radius of the emitting region. Both $p$ and $\overline{r}$ are of order unity, meaning that $x$ is generally not constant and depends on $\dot{m}$ through $l$. However, $x$ is often set to unity for simplicity, producing the classical form of the beaming factor:
\begin{equation}
b = \left\{
    \begin{aligned}
        & 1, \quad & \dot{m} \leq 8.5 \\
        & \left(\frac{8.5}{\dot{m}}\right)^2, \quad & \dot{m} > 8.5
    \end{aligned}
\right.
\end{equation}
Here, the critical accretion rate $\dot{m}_{\rm crit} = 8.5$ is defined by the condition $b = 1$, corresponding physically to the threshold above which beaming effects become significant.

When the $\dot{m}$-dependence of $x$ is properly taken into account, we obtain a log-modified prescription that modifies the behavior of $b$ at high accretion rates:
\begin{equation}
b = \left\{
    \begin{aligned}
        & 1, \quad & \dot{m} \leq 38.5 \\
        & {\left(\frac{8.5}{\dot{m}}\right)^2 (1 + \ln \dot{m})^2}, 
        \quad & {\dot{m} > 38.5,}
    \end{aligned}
\right.
\end{equation}

In this case, the critical accretion rate $\dot{m}_{\rm crit} = 38.5$ is again determined by the condition $b = 1$. The higher value of $\dot{m}_{\rm crit}$ indicates that, compared to the classical model, the log-modified model reflects a weaker degree of collimation, implying that a higher accretion rate is required for beaming to become significant. This behavior is shown in the middle panel of Figure \ref{fig:figure1}. For $\dot{m} > 38.5$, the classical and log-modified prescriptions differ by roughly an order of magnitude, with the discrepancy increasing at higher $\dot{m}$.

The observed luminosity $L_{\rm {iso}}$, estimated under the assumption of isotropic emission, can be expressed as:
\begin{equation}
L_{\text{iso}} = \frac{L_{\text{int}}}{b}.
\end{equation}
This gives the following piecewise forms for $L_{\text{iso}}$ in the two models.

For the classical model:
\begin{equation}
L_{\text{iso}} = L_{\rm Edd} \cdot \left\{
    \begin{aligned}
        & \dot{m}, \quad & \dot{m} \leq 1 \\
        & 1 + \ln \dot{m}, \quad & 1 < \dot{m} \leq 8.5 \\
        & \left(\frac{\dot{m}}{8.5}\right)^2 (1 + \ln \dot{m}), 
        \quad & \dot{m} > 8.5,\ {\rm BH}\\
        & \left(\frac{\dot{m}}{8.5}\right)^2 
        \max\!\left(1 + \ln \dot{m},\,0.125\dot{m}\right), 
        \quad & \dot{m} > 8.5,\ {\rm NS}
    \end{aligned}
\right.
\end{equation}

For the log-modified model:
\begin{equation}
L_{\text{iso}} = L_{\rm Edd} \cdot \left\{
    \begin{aligned}
        & \dot{m}, \quad & \dot{m} \leq 1 \\
        & 1 + \ln \dot{m}, \quad & 1 < \dot{m} \leq 38.5 \\
        & \left(\frac{\dot{m}}{8.5}\right)^2 (1 + \ln \dot{m})^{-1}, 
        \quad & \dot{m} > 38.5,\ {\rm BH}\\
        & \left(\frac{\dot{m}}{8.5}\right)^2 (1 + \ln \dot{m})^{-2} \max\!\left(1 + \ln \dot{m},\,0.125\dot{m}\right), 
        \quad & \dot{m} > 38.5,\ {\rm NS}
    \end{aligned}
\right.
\end{equation}
Here, the factor $(1 + \ln \dot{m})^{-1}$ and $(1 + \ln \dot{m})^{-2}$ in the high-$\dot{m}$ regime arises from the inverse dependence of $b$ on $l^2$, reflecting the more gradual increase of collimation with accretion rate in the log-modified scenario. The resulting $L_{\rm iso}$–$\dot{m}$ relations for both the classical and log-modified models are shown in the bottom panel of Figure \ref{fig:figure1}. At low accretion rates ($\dot{m} < 8.5$), the two prescriptions have identical results, while at high accretion rates, the log-modified model predicts significantly lower observed luminosities due to the weaker beaming effect, with the difference between the two curves growing rapidly as $\dot{m}$ increases, consistent with the corresponding behavior of $b(\dot{m})$.

Finally, since the beaming factor is given by $b = \Omega / 4\pi$, it can be directly interpreted as the probability that a ULX is observable from a random viewing angle. In our simulations, this selection effect is introduced by generating a random number from a uniform distribution in the range $(0,1)$ for each source: if the number is smaller than $b$, the source is considered detectable; otherwise, it is regarded as invisible.

\subsection{Binary population synthesis}
We employ the binary population synthesis (BPS) code originally developed by \citet{Hurley2000, Hurley2002} to simulate the evolution of $5\times 10^{5}$ binary systems at both solar and sub-solar metallicities ($Z=0.016$ and $0.0016$). The initial binary parameters are generated as follows. The primary mass $M_1$ is sampled from the \citet{Kroupa1993} initial mass function within the range $0.08$ – $60~{\rm M_{\odot}}$, while the mass ratio $q=M_2/M_1$ is drawn from a uniform distribution between 0 and 1 \citep{Kobulnicky2007}. Orbital separations $a$ are chosen from a logarithmically flat distribution between $3$ and $10^{4}~{\rm R_{\odot}}$ \citep{Abt1983}, and all binaries are assumed to start in circular orbits.

We evolve binaries starting from two zero-age main-sequence stars. As the system evolves, the more massive primary star undergoes a supernova explosion first. If the CO core mass at the time of explosion exceeds 7 $\rm M_{\odot}$, we assume the star collapse into a BH, whereas smaller cores result in NSs \citep{Hurley2000}. The system then continues to evolve until the secondary star overflows its Roche-lobe (RL) and transfers mass to the BH/NS.
During this later phase, we record key evolutionary parameters, including the mass-loss rate $\dot{M}_2$, the timestep $\Delta t$, and the RL filling status of the donor star. 
We consider mass loss from stellar winds and stable RL overflow, excluding the common-envelope phase. Roche-lobe overflow (RLOF) occurs when the donor’s radius exceeds its Roche-lobe radius, calculated using the \citet{Eggleton1983} approximation. The distinction between stable and unstable mass transfer, as well as the mass-transfer rate during RLOF, follows the criteria described in \citet[Sec.~2.6]{Hurley2002}. Stellar winds are treated using a combination of prescriptions: \citet{Nieuwenhuijzen1990} for massive stars throughout their evolution, Reimers mass loss for low- and intermediate-mass giants \citep{Reimers1975} , and \citet{Vassiliadis1993} for AGB stars. All other physical assumptions adopt the default settings of the BPS code.

To ensure that all potential ULX progenitors are captured, we select evolutionary stages with $\dot{M}_2 > 10^{17}~\rm{g~s^{-1}}$, corresponding to approximately $10\%$ of the Eddington accretion rate for a solar-mass compact object. For RLOF systems, we directly adopt the mass transfer rate from the donor as the accretion rate, i.e., $\dot{M} = \dot{M}_2$. For wind accretion, we consider the spherically symmetric gravitational capture described by the Bondi–Hoyle accretion, where $\dot{M}$ can be written as \citep{Bondi1944}:
\begin{equation}
    \dot{M}=\pi R_{\rm G}^{2} \rho v_{\rm rel}=\frac{\pi R_{\rm G}^{2}\dot{M}_{2}v_{\rm rel}}{4\pi a^{2}v_{\rm w}},
\end{equation}
where $\rho$ is the stellar wind density, $v_{\rm rel} = \sqrt{v_{\rm w}^2 + v_{\rm orb}^2}$ is the relative velocity between the accretor and the wind, $v_{\rm w}$ is the wind velocity, $v_{\rm orb}$ is the orbital velocity of the accretor, $a$ is the binary separation, and $R_{\rm G} = 2GM_1/v_{\rm rel}^2$ denotes the gravitational capture radius of the compact star with mass $M_1$. The stellar wind velocity of the donor star follows \citet{Castor1975}:
\begin{equation}
v_{\rm w} = \alpha v_{\rm esc} \left(1 - \frac{R_2}{a}\right)^{\beta},
\end{equation}
where $v_{\rm esc} = \sqrt{2GM_2/R_2}$ is the escape velocity from the donor’s surface, with $M_2$ and $R_2$ denoting the donor’s mass and radius, respectively. The coefficient $\alpha$ and the power-law index $\beta$ are set to 1 and 0.8, respectively \citep{Waters1989}. All the above quantities are directly obtained from the BPS output. With these prescriptions, and using the parameters obtained from the BPS together with the beaming factor $b(\dot{m})$, the intrinsic luminosity $L_{\rm int}(\dot{m})$, and the isotropic luminosity $L_{\rm iso}(\dot{m})$ calculated in Section~\ref{subsec:2.1}, we derive the statistical properties of the ULX population and compare them with observational constraints.

\begin{figure}[ht!]
\centering
\includegraphics[height=12cm]{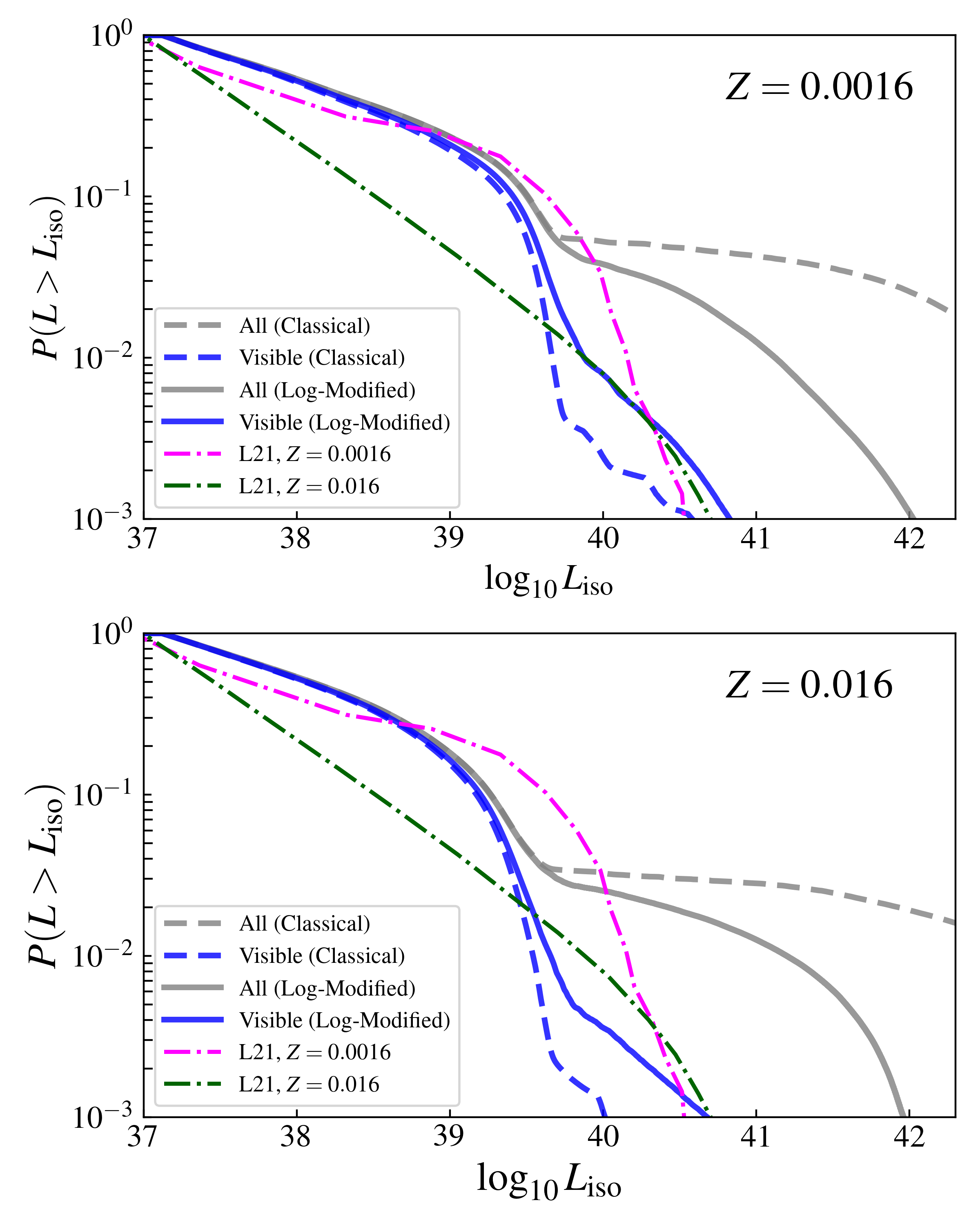}  
\caption{
Simulated and observed XLFs at two metallicities: $Z = 0.0016$ (top) and $Z = 0.016$ (bottom). Simulated results are shown as gray curves for the total population and blue curves for the cumulative visible population, with dashed and solid lines indicating the classical and log-modified models, respectively. Observed XLFs are overplotted as thin magenta ($Z = 0.0016$) and green ($Z = 0.016$) lines.
}
\label{fig:figure2}
\end{figure}

\begin{figure}[ht!]
\centering
\includegraphics[height=12cm]{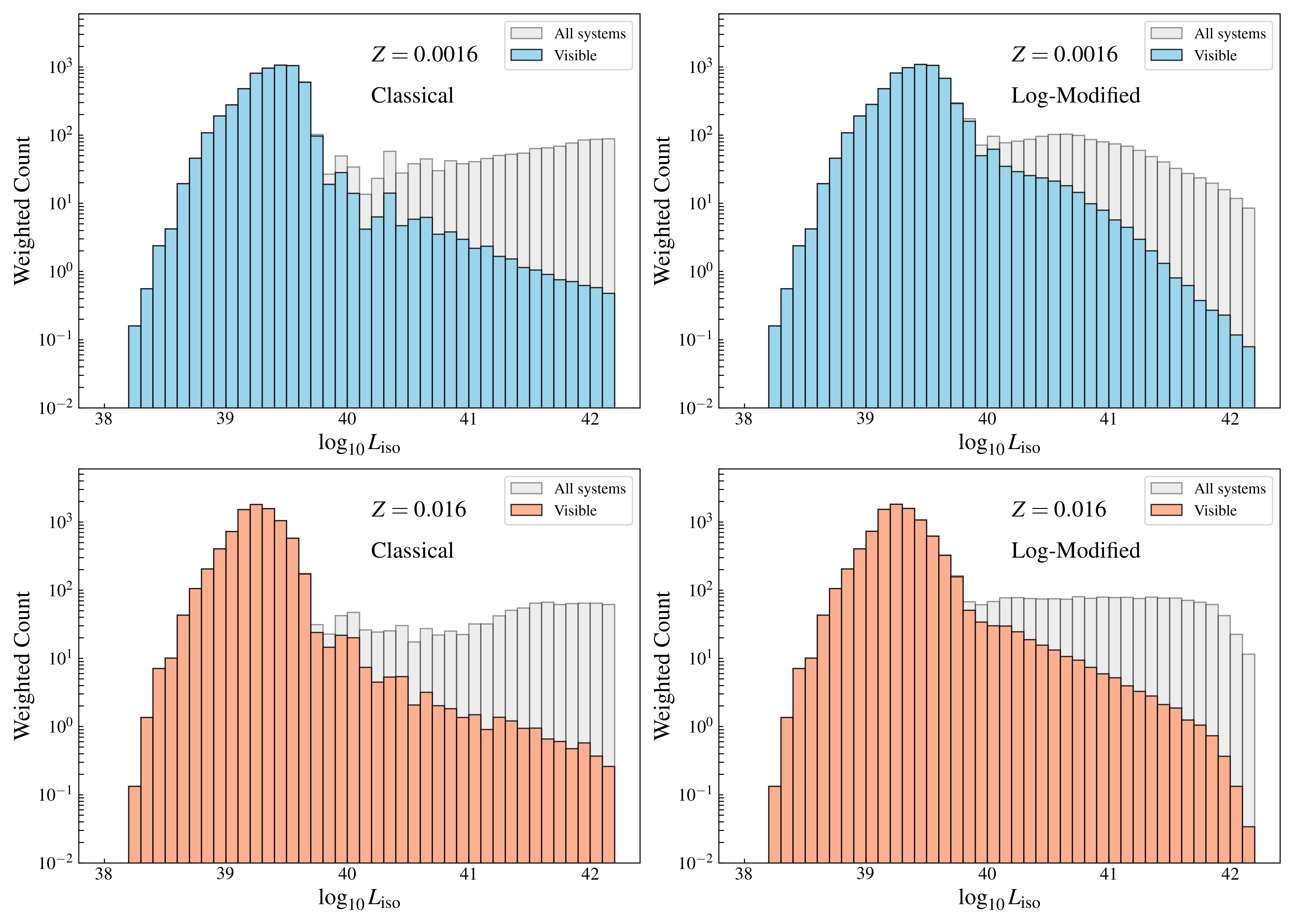}  
\caption{
Probability distribution histograms of ULX counts as a function of isotropic luminosity. The vertical axis shows time-weighted counts per luminosity bin. Gray bars denote the total number of systems, while blue (classical model) and salmon (log-modified model) bars indicate the visible subsets. The top panels correspond to $Z = 0.0016$, and the bottom panels correspond to $Z = 0.016$, respectively.
}
\label{fig:figure3}
\end{figure}

\begin{figure}[ht!]
\centering
\includegraphics[height=8cm]{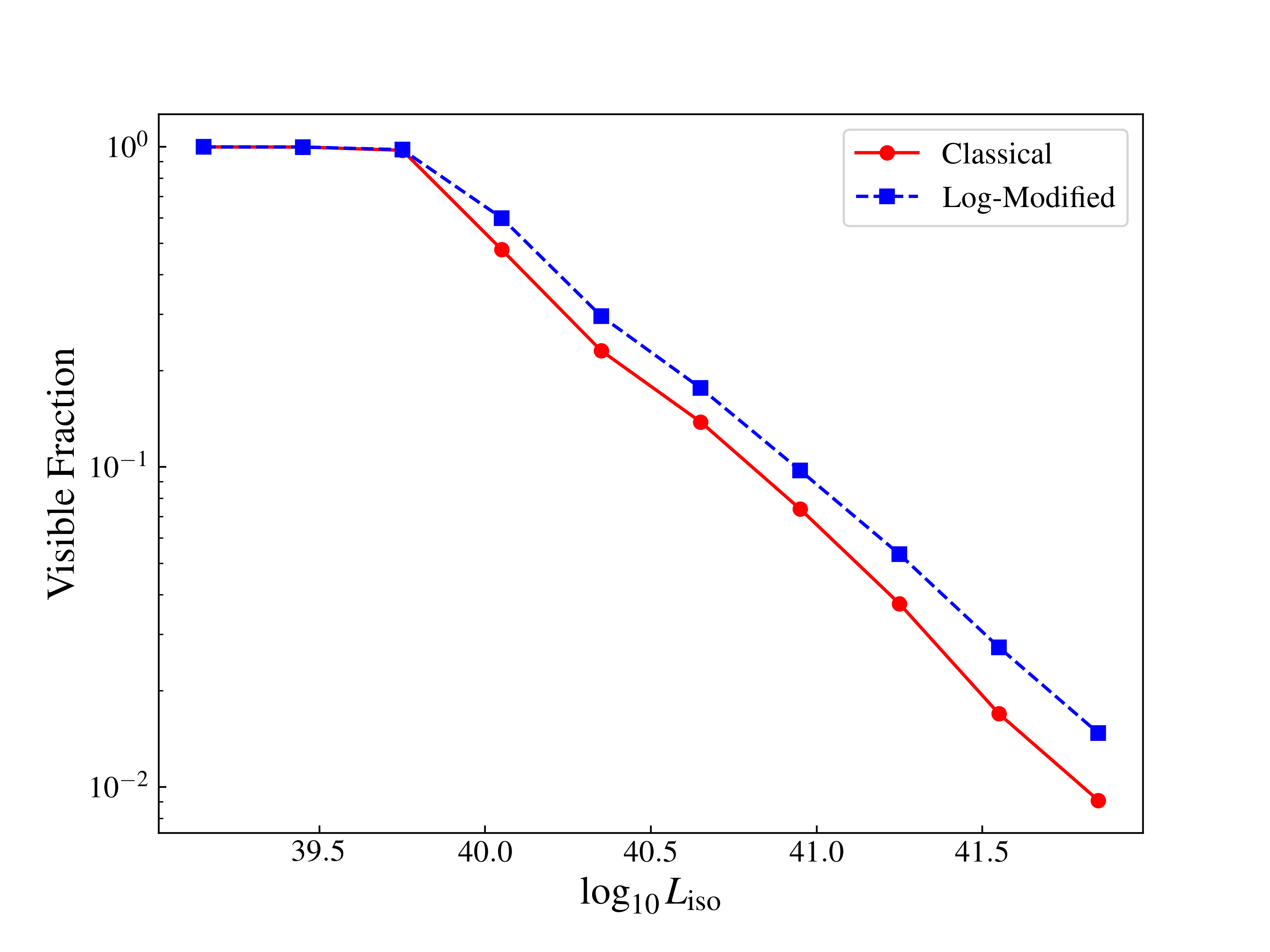}  
\caption{
Observable fraction of ULXs as a function of luminosity. The horizontal axis shows $\log_{10}L_{\rm iso}$, and the vertical axis shows the visible fraction, defined as the ratio of visible systems to the total number of systems in each luminosity bin. The red circles connected by a solid line represent the classical model, while the blue squares connected by a dashed line represent the log-modified model. Since metallicity has only a minor impact on this ratio, only the $Z = 0.016$ case is shown.
}
\label{fig:figure4}
\end{figure}

\begin{figure}[ht!]
\centering
\includegraphics[height=12cm]{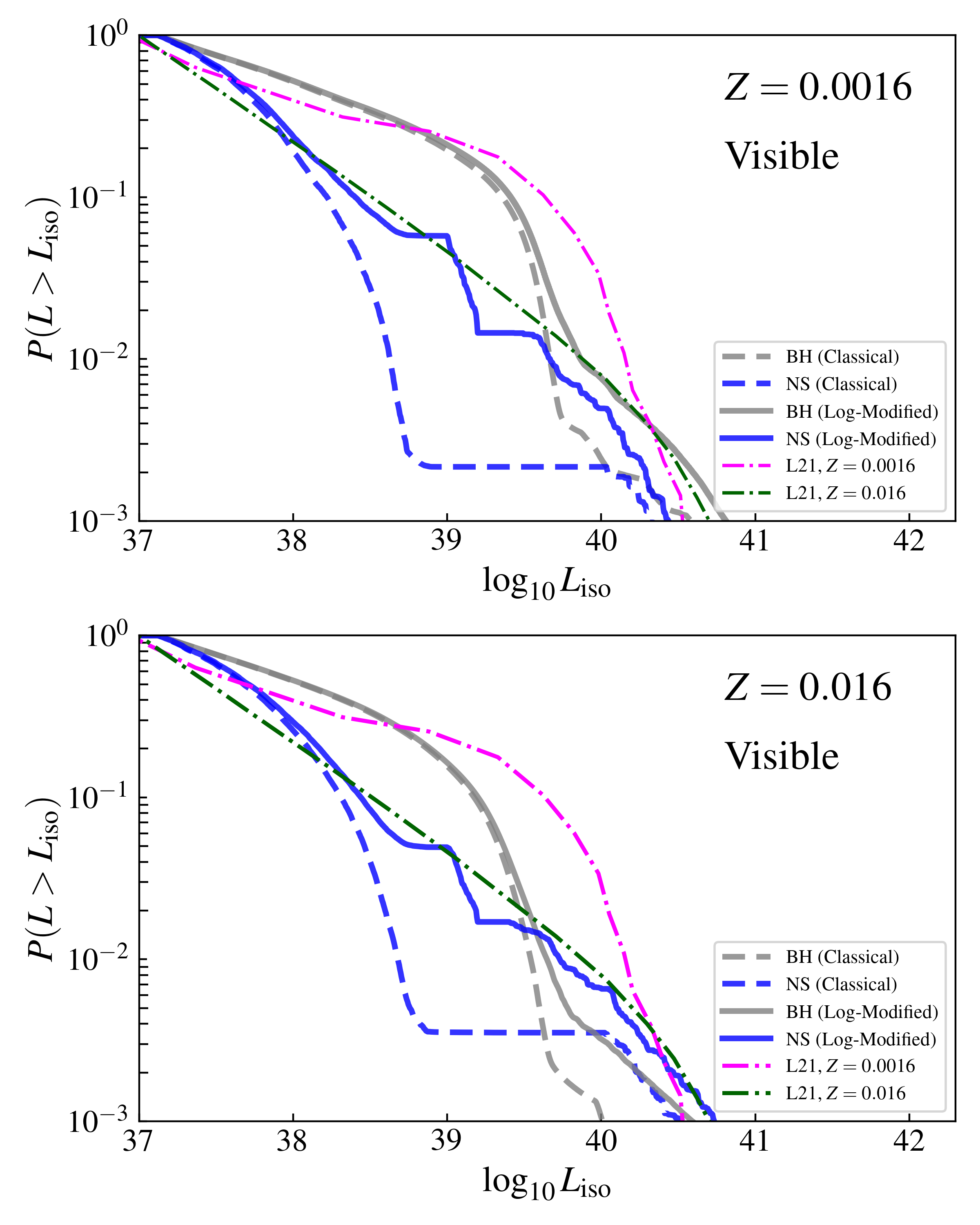}  
\caption{
Visible XLFs of BH and NS ULXs at two metallicities. The upper and lower panels correspond to $Z = 0.0016$ and $Z = 0.016$, respectively. Dashed and solid lines denote the classical and log-modified models, while blue and gray curves represent NSs and BHs. For comparison, the thin magenta dotted line shows the observed XLF at $Z = 0.0016$, and the thin green dashed line shows the observed XLF at $Z = 0.016$.
}
\label{fig:figure5}
\end{figure}

\section{Results} \label{sec:Results}

The XLF describes  the abundance of sources with luminosities surpassing a specified threshold, offering a statistical overview of population luminosity. To emulate this characteristic in our simulations, we construct the XLF using time-weighted cumulative distributions of isotropic luminosity $L_{\rm iso}$  across evolutionary stages within the initial 100 Myr. Each luminosity bin is weighted according to the duration spent at that luminosity level, approximating the cumulative contribution from a 100 Myr star-forming epoch and implicitly establishing a lower bound on donor mass.

In \citet{Lehmer2021} (hereafter L21), the authors investigated the metallicity dependence of the high-mass X-ray binary (HMXB) XLF in nearby galaxies within a distance of $D \lesssim 30$ Mpc. Adopting the aforementioned methodology, we compute XLFs at $Z=0.016\ (\sim  Z_{\odot})$ and $Z=0.0016\ (\sim 0.1~Z_{\odot})$, specifically chosen to facilitate direct comparison with L21's observational findings. For a shape-only comparison, both observed and simulated XLFs are normalized such that $P(L_{\rm iso} >10^{37}\rm{erg~s^{-1}}) = 1$, facilitating a direct evaluation of the high-luminosity slope.

The observational and simulated XLFs are shown in Figure \ref{fig:figure2}, where the upper and lower panels correspond to $Z = 0.0016$ and $Z = 0.016$, respectively. In each panel, the gray dashed and solid lines represent the total number of systems for the classical and log-modified models, while the blue dashed and solid lines denote the cumulative number of visible systems under the same two models. For reference, the thin magenta dash-dotted  line corresponds to the observation at $Z = 0.0016$ from L21, and the thin green line shows the result for $Z = 0.016$ from the same study. The general trend indicates that lower metallicity yields a higher proportion of sources at the high-luminosity end, a pattern consistently observed in both simulations and observations.

When selection effects are not applied (corresponding to the two gray lines in the figure), the classical model predicts a greater number of sources in $L_{\rm iso}>10^{40}$ regime compared to the log-modified model. This discrepancy arises from the beaming prescription: for a given accretion rate $\dot{m}$, the classical model yields smaller beaming factors $b$, enhancing beaming and shifting  more systems into the bright end of the distribution. Since $b$ also influences observability, its rapid decline with increasing $\dot{m}$ subsequently reduces the visible population in the classical scenario. Consequently, although the classical model generates more high-luminosity ULXs overall, its visible XLF falls below that of the log-modified model.

It is important to note that BPS calculation inherently simplifies many physical processes; therefore, a mismatch between the predicted and observed ULX populations is not unexpected. Critical parameters that shape the ULX phase, including the mass-loss rate and the fraction of transferred material during RLOF, cannot be reconstructed with high accuracy in population synthesis. We employ the logarithmic formula from the SS model to describe super-Eddington accretion, yet this approach overlooks the effects of advection — a significant limitation when applying the formula to ULXs. Additionally, the potentially strong impact of the NS magnetic field on the ULX population \citep[e.g.,][]{Kuranov2020,Kovlakas2025} is not included in BPS. The comparison is further affected by observational incompleteness and the strong dependence of the inferred luminosity on the beaming factor. Therefore, our preference for the two beaming formulae is based on their ability to reproduce the observed trends, rather than on an exact match to the observed ULX numbers.

We perform a Kolmogorov–Smirnov test on the simulated and observed XLFs to evaluate how well the two models reproduce the data. For $Z=0.0016$ and $Z=0.016$, the p-values for the classical model are 0.21 and 0.72, respectively, while the log-modified model produces p-values of 0.95 for both metallicities. Higher p-values indicate a closer agreement between the model and the observations. For both metallicities, the log-modified model achieves consistently high p-values, suggesting that it reproduces the overall ULX distribution well. The classical model performs less favorably, particularly at $Z=0.0016$, where the moderate p-value points to a noticeable deviation from the data. Although the classical model shows improved agreement at $Z=0.016$, its performance remains below that of the log-modified model. These results indicate that the log-modified prescription provides a better description of the ULX population, capturing both the luminosity distribution and its metallicity dependence more effectively than the classical model.

Comparison with L21's observational data reveals that, without accounting for selection effects, both models overestimate the number of sources at the bright end ($L_{\rm iso} \gtrsim 10^{40}~{\rm erg~s^{-1}}$). These luminous ULXs are associated with high accretion rates and extremely small beaming factors ($b \lesssim 0.01$). Thus, applying visibility selection criteria significantly diminishes their numbers. The log-modified model retains a larger fraction of bright systems, yielding a high-luminosity slope that more closely aligns with observed trends.

This pattern is further elucidated  in Figure \ref{fig:figure3}, which presents probability distribution histograms of ULX counts across  various luminosities. Gray bars indicate the weighted total number of systems, while blue or salmon-colored bars denote the visible subset. Below $10^{40}~{\rm erg~s^{-1}}$, nearly all systems are visible due to negligible beaming. However, at higher luminosities, the visible population in the classical model (left panels) is consistently smaller than in the log-modified model (right panels), a disparity that persists up to $10^{42}~{\rm erg~s^{-1}}$. In the log-modified  scenario, the inclusion of the $(1+\ln\dot{m})^2$ term mitigates the beaming effect, preventing  excessively small $b$ values ($<10^{-3}$). This adjustment  curtails the extreme luminous tail while maintaining lower-luminosity visibility, resulting in a more equitable distribution of observable ULXs and enhanced agreement with observed XLFs. 

By calculating the ratio of visible systems to the total number of systems in each luminosity bin, we derive the observable fraction depicted in Figure \ref{fig:figure4}. This fraction can be interpreted as the time-weighted average of the beaming factor $b$ across the population. The results demonstrate that the classical model consistently produces lower observable fractions than the log-modified model, typically reaching only $60$–$80\%$ of the latter. Consequently, the log-modified model produces a more pronounced high-luminosity tail in the XLF by preserving more luminous sources after selection effects. In both cases, the fraction rapidly declines from $\sim 0.5$ at $10^{40}~{\rm erg~s^{-1}}$ to $\sim 0.01$ at $10^{42}~{\rm erg~s^{-1}}$. This trend underscores that the beaming  selection effect significantly diminishes the visible population at the high-luminosity tail.

In Figure \ref{fig:figure5}, we dissect the XLFs of BHs and NSs to  evaluate their contributions to the overall ULX population, without considering the impact of NS magnetic fields on Eddington luminosity. All the curves represent the visible subsets. The BH ULX population predominantly shapes the overall XLF. However, a notable distinction emerges within the NS ULX population: the log-modified model forecasts a substantially larger fraction of observable NS systems above $10^{39}~{\rm erg~s^{-1}}$, exceeding that of the classical model by more than an order of magnitude. To achieve $L_{\rm iso} \sim 10^{39}~{\rm erg~s^{-1}}$, a typical NS ULX requires $\dot{m}\sim10$ and $b\sim0.6$ in the classical model, whereas the same luminosity necessitates $\dot{m}\sim40$ and $b\sim0.9$ in the log-modified model. Given that weaker beaming enhances visibility, the log-modified prescription augments the contribution of NS ULXs, particularly at intermediate luminosities. This increased NS visibility may help bridge the gap between theoretical predictions and observed NS ULX fractions.

\section{Conclusions} \label{sec:Conclusion}
In this work, we  explore the impact of various formulations of the beaming factor $b(\dot{m})$  on the predicted properties of ULX populations, employing BPS simulations at both solar and sub-solar metallicities. By comparing the widely adopted classical model with a log-modified model that incorporates the $(1+\ln\dot{m})^2$ dependence at high accretion rates, we evaluate their effects on overall and observable luminosity distributions, as well as on the relative contributions of BHs and NSs to the ULX population.

Our results demonstrate that the selection of beaming prescription significantly shapes the population properties. The classical model yields a larger total number of high-luminosity ULXs due to intensified beaming, which amplifies isotropic luminosity. However, this enhancement is offset by reduced visibility: the sharp decline in the beaming factor with accretion rate diminishes the fraction of observable systems, resulting in an XLF that falls below that of the log-modified model. Conversely, the log-modified prescription mitigates beaming in the intermediate accretion regime, thereby increasing the observable fraction of systems and generating a high-luminosity slope that aligns more closely with observed XLFs from nearby galaxies.

A notable consequence of these disparities arises for NS ULXs. The classical model forecasts a limited number of visible NS systems, as achieving luminosities above $10^{39}~\rm erg~s^{-1}$ necessitates extreme accretion rates coupled  with strong beaming, significantly reducing their detectability. In contrast, the log-modified model permits weaker collimation at comparable luminosities, thereby enhancing the observable fraction of NS ULXs by nearly an order of magnitude compared to the classical prescription.

Relative to the classical model, the log-modified model suggests a more moderate beaming scenario, where extreme values of small $b$ and large $L_{\rm iso}$ are less probable. This may offer a more realistic representation of geometrical collimation and accretion physics in the super-Eddington regime. Despite these advancements, our study also acknowledges certain limitations. Both formulations, after accounting for selection effects, tend to marginally underestimate the number of sources at $L_{\rm iso}>10^{40}$ when compared with observations. This discrepancy may stem from uncertainties in the simulations, such as accretion disk instabilities and magnetospheric truncation, as well as potential deviations of the beaming factor from purely theoretical values in real-world scenarios. Further research is warranted to address these uncertainties and refine our comprehension of the ULX population in extreme environments.

In conclusion, our study demonstrates that the log-modified beaming prescription offers a superior fit to observed ULX luminosity functions and provides a more credible explanation for the substantial presence of pulsing ULXs. These insights underscore the significance of employing realistic beaming laws in ULX population studies. The choice of beaming factor is pivotal for enhancing theoretical predictions and deepening our understanding of the compact object populations that drive ULXs.
\bibliography{ref}{}

\begin{thebibliography}{}
\expandafter\ifx\csname natexlab\endcsname\relax\def\natexlab#1{#1}\fi
\providecommand{\url}[1]{\href{#1}{#1}}
\providecommand{\dodoi}[1]{doi:~\href{http://doi.org/#1}{\nolinkurl{#1}}}
\providecommand{\doeprint}[1]{\href{http://ascl.net/#1}{\nolinkurl{http://ascl.net/#1}}}
\providecommand{\doarXiv}[1]{\href{https://arxiv.org/abs/#1}{\nolinkurl{https://arxiv.org/abs/#1}}}

\bibitem[{{Abarca} {et~al.}(2021){Abarca}, {Parfrey}, \&
  {Klu{\'z}niak}}]{Abarca2021}
{Abarca}, D., {Parfrey}, K., \& {Klu{\'z}niak}, W. 2021, \apjl, 917, L31,
  \dodoi{10.3847/2041-8213/ac1859}

\bibitem[{{Abt}(1983)}]{Abt1983}
{Abt}, H.~A. 1983, \araa, 21, 343, \dodoi{10.1146/annurev.aa.21.090183.002015}

\bibitem[{{Bachetti} {et~al.}(2014){Bachetti}, {Harrison}, {Walton},
  {Grefenstette}, {Chakrabarty}, {F{\"u}rst}, {Barret}, {Beloborodov}, {Boggs},
  {Christensen}, {Craig}, {Fabian}, {Hailey}, {Hornschemeier}, {Kaspi},
  {Kulkarni}, {Maccarone}, {Miller}, {Rana}, {Stern}, {Tendulkar}, {Tomsick},
  {Webb}, \& {Zhang}}]{Bachetti2014}
{Bachetti}, M., {Harrison}, F.~A., {Walton}, D.~J., {et~al.} 2014, \nat, 514,
  202, \dodoi{10.1038/nature13791}

\bibitem[{{Baschek} {et~al.}(1975){Baschek}, {Kegel}, \&
  {Traving}}]{Reimers1975}
{Baschek}, B., {Kegel}, W.~H., \& {Traving}, G., eds. 1975, {Problems in
  Stellar Atmospheres and Envelopes} (Springer), 229.
\newblock \url{https://link.springer.com/book/10.1007/978-3-642-80919-4}

\bibitem[{{Binder} {et~al.}(2018){Binder}, {Levesque}, \&
  {Dorn-Wallenstein}}]{Binder2018}
{Binder}, B., {Levesque}, E.~M., \& {Dorn-Wallenstein}, T. 2018, \apj, 863,
  141, \dodoi{10.3847/1538-4357/aad3bd}

\bibitem[{{Bondi} \& {Hoyle}(1944)}]{Bondi1944}
{Bondi}, H., \& {Hoyle}, F. 1944, \mnras, 104, 273,
  \dodoi{10.1093/mnras/104.5.273}

\bibitem[{{Castor} {et~al.}(1975){Castor}, {Abbott}, \& {Klein}}]{Castor1975}
{Castor}, J.~I., {Abbott}, D.~C., \& {Klein}, R.~I. 1975, \apj, 195, 157,
  \dodoi{10.1086/153315}

\bibitem[{{Chashkina} {et~al.}(2019){Chashkina}, {Lipunova}, {Abolmasov}, \&
  {Poutanen}}]{Chashkina2019}
{Chashkina}, A., {Lipunova}, G., {Abolmasov}, P., \& {Poutanen}, J. 2019, \aap,
  626, A18, \dodoi{10.1051/0004-6361/201834414}

\bibitem[{{Colbert} \& {Mushotzky}(1999)}]{Colbert1999}
{Colbert}, E. J.~M., \& {Mushotzky}, R.~F. 1999, \apj, 519, 89,
  \dodoi{10.1086/307356}

\bibitem[{{Eggleton}(1983)}]{Eggleton1983}
{Eggleton}, P.~P. 1983, \apj, 268, 368, \dodoi{10.1086/160960}

\bibitem[{{Fabbiano}(1989)}]{Fabbiano1989}
{Fabbiano}, G. 1989, \araa, 27, 87, \dodoi{10.1146/annurev.aa.27.090189.000511}

\bibitem[{{Hurley} {et~al.}(2000){Hurley}, {Pols}, \& {Tout}}]{Hurley2000}
{Hurley}, J.~R., {Pols}, O.~R., \& {Tout}, C.~A. 2000, \mnras, 315, 543,
  \dodoi{10.1046/j.1365-8711.2000.03426.x}

\bibitem[{{Hurley} {et~al.}(2002){Hurley}, {Tout}, \& {Pols}}]{Hurley2002}
{Hurley}, J.~R., {Tout}, C.~A., \& {Pols}, O.~R. 2002, \mnras, 329, 897,
  \dodoi{10.1046/j.1365-8711.2002.05038.x}

\bibitem[{{Kaaret} {et~al.}(2017){Kaaret}, {Feng}, \& {Roberts}}]{Kaaret2017}
{Kaaret}, P., {Feng}, H., \& {Roberts}, T.~P. 2017, \araa, 55, 303,
  \dodoi{10.1146/annurev-astro-091916-055259}

\bibitem[{{Kajava} \& {Poutanen}(2008)}]{Kajava2008}
{Kajava}, J. J.~E., \& {Poutanen}, J. 2008, in American Institute of Physics
  Conference Series, Vol. 1054, Cool Discs, Hot Flows: The Varying Faces of
  Accreting Compact Objects, ed. M.~{Axelsson}, 39--47,
  \dodoi{10.1063/1.3002507}

\bibitem[{{Kayanikhoo} {et~al.}(2025){Kayanikhoo}, {Klu{\'z}niak}, {Abarca}, \&
  {Cemeljic}}]{Kayanikhoo2025}
{Kayanikhoo}, F., {Klu{\'z}niak}, W., {Abarca}, D., \& {Cemeljic}, M. 2025,
  arXiv e-prints, arXiv:2508.02563, \dodoi{10.48550/arXiv.2508.02563}

\bibitem[{{King} \& {Lasota}(2019)}]{King2019}
{King}, A., \& {Lasota}, J.-P. 2019, \mnras, 485, 3588,
  \dodoi{10.1093/mnras/stz720}

\bibitem[{{King} \& {Lasota}(2024)}]{King2024}
---. 2024, \aap, 682, L22, \dodoi{10.1051/0004-6361/202349002}

\bibitem[{{King} {et~al.}(2017){King}, {Lasota}, \& {Klu{\'z}niak}}]{King2017}
{King}, A., {Lasota}, J.-P., \& {Klu{\'z}niak}, W. 2017, \mnras, 468, L59,
  \dodoi{10.1093/mnrasl/slx020}

\bibitem[{{King}(2009)}]{King2009}
{King}, A.~R. 2009, \mnras, 393, L41, \dodoi{10.1111/j.1745-3933.2008.00594.x}

\bibitem[{{King} {et~al.}(2001){King}, {Davies}, {Ward}, {Fabbiano}, \&
  {Elvis}}]{King2001}
{King}, A.~R., {Davies}, M.~B., {Ward}, M.~J., {Fabbiano}, G., \& {Elvis}, M.
  2001, \apjl, 552, L109, \dodoi{10.1086/320343}

\bibitem[{{Kobulnicky} \& {Fryer}(2007)}]{Kobulnicky2007}
{Kobulnicky}, H.~A., \& {Fryer}, C.~L. 2007, \apj, 670, 747,
  \dodoi{10.1086/522073}

\bibitem[{{K{\"o}rding} {et~al.}(2002){K{\"o}rding}, {Falcke}, \&
  {Markoff}}]{Kording2002}
{K{\"o}rding}, E., {Falcke}, H., \& {Markoff}, S. 2002, \aap, 382, L13,
  \dodoi{10.1051/0004-6361:20011776}

\bibitem[{{Kovlakas} {et~al.}(2025){Kovlakas}, {Misra}, {Amato}, \& {Luca
  Israel}}]{Kovlakas2025}
{Kovlakas}, K., {Misra}, D., {Amato}, R., \& {Luca Israel}, G. 2025, \aap, 694,
  L9, \dodoi{10.1051/0004-6361/202453274}

\bibitem[{{Kroupa} {et~al.}(1993){Kroupa}, {Tout}, \& {Gilmore}}]{Kroupa1993}
{Kroupa}, P., {Tout}, C.~A., \& {Gilmore}, G. 1993, \mnras, 262, 545,
  \dodoi{10.1093/mnras/262.3.545}

\bibitem[{{Kuranov} {et~al.}(2020){Kuranov}, {Postnov}, \&
  {Yungelson}}]{Kuranov2020}
{Kuranov}, A.~G., {Postnov}, K.~A., \& {Yungelson}, L.~R. 2020, Astronomy
  Letters, 46, 658, \dodoi{10.1134/S1063773720100084}

\bibitem[{{Kuranov} {et~al.}(2021){Kuranov}, {Postnov}, \&
  {Yungelson}}]{Kuranov2021}
---. 2021, Astronomy Letters, 47, 831, \dodoi{10.1134/S1063773721120021}

\bibitem[{{Lehmer} {et~al.}(2021){Lehmer}, {Eufrasio}, {Basu-Zych}, {Doore},
  {Fragos}, {Garofali}, {Kovlakas}, {Williams}, {Zezas}, \&
  {Santana-Silva}}]{Lehmer2021}
{Lehmer}, B.~D., {Eufrasio}, R.~T., {Basu-Zych}, A., {et~al.} 2021, \apj, 907,
  17, \dodoi{10.3847/1538-4357/abcec1}

\bibitem[{{Miller} {et~al.}(2003){Miller}, {Fabbiano}, {Miller}, \&
  {Fabian}}]{Miller2003}
{Miller}, J.~M., {Fabbiano}, G., {Miller}, M.~C., \& {Fabian}, A.~C. 2003,
  \apjl, 585, L37, \dodoi{10.1086/368373}

\bibitem[{{Mushtukov} {et~al.}(2021){Mushtukov}, {Portegies Zwart},
  {Tsygankov}, {Nagirner}, \& {Poutanen}}]{Mushtukov2021}
{Mushtukov}, A.~A., {Portegies Zwart}, S., {Tsygankov}, S.~S., {Nagirner},
  D.~I., \& {Poutanen}, J. 2021, \mnras, 501, 2424,
  \dodoi{10.1093/mnras/staa3809}

\bibitem[{{Mushtukov} {et~al.}(2015){Mushtukov}, {Suleimanov}, {Tsygankov}, \&
  {Poutanen}}]{Mushtukov2015}
{Mushtukov}, A.~A., {Suleimanov}, V.~F., {Tsygankov}, S.~S., \& {Poutanen}, J.
  2015, \mnras, 454, 2539, \dodoi{10.1093/mnras/stv2087}

\bibitem[{{Nieuwenhuijzen} \& {de Jager}(1990)}]{Nieuwenhuijzen1990}
{Nieuwenhuijzen}, H., \& {de Jager}, C. 1990, \aap, 231, 134

\bibitem[{{Poutanen} {et~al.}(2007){Poutanen}, {Lipunova}, {Fabrika},
  {Butkevich}, \& {Abolmasov}}]{Poutanen2007}
{Poutanen}, J., {Lipunova}, G., {Fabrika}, S., {Butkevich}, A.~G., \&
  {Abolmasov}, P. 2007, \mnras, 377, 1187,
  \dodoi{10.1111/j.1365-2966.2007.11668.x}

\bibitem[{{Shakura} \& {Sunyaev}(1973)}]{Shakura1973}
{Shakura}, N.~I., \& {Sunyaev}, R.~A. 1973, \aap, 24, 337

\bibitem[{{Shao} \& {Li}(2015)}]{Shao2015}
{Shao}, Y., \& {Li}, X.-D. 2015, \apj, 802, 131,
  \dodoi{10.1088/0004-637X/802/2/131}

\bibitem[{{Shao} {et~al.}(2019){Shao}, {Li}, \& {Dai}}]{Shao2019}
{Shao}, Y., {Li}, X.-D., \& {Dai}, Z.-G. 2019, \apj, 886, 118,
  \dodoi{10.3847/1538-4357/ab4d50}

\bibitem[{{Sutton} {et~al.}(2013){Sutton}, {Roberts}, \&
  {Middleton}}]{Sutton2013}
{Sutton}, A.~D., {Roberts}, T.~P., \& {Middleton}, M.~J. 2013, \mnras, 435,
  1758, \dodoi{10.1093/mnras/stt1419}

\bibitem[{{Vassiliadis} \& {Wood}(1993)}]{Vassiliadis1993}
{Vassiliadis}, E., \& {Wood}, P.~R. 1993, \apj, 413, 641,
  \dodoi{10.1086/173033}

\bibitem[{{Waters} \& {van Kerkwijk}(1989)}]{Waters1989}
{Waters}, L. B. F.~M., \& {van Kerkwijk}, M.~H. 1989, \aap, 223, 196

\bibitem[{{Wiktorowicz} {et~al.}(2021){Wiktorowicz}, {Lasota}, {Belczynski},
  {Lu}, {Liu}, \& {I{\l}kiewicz}}]{Wiktorowicz2021}
{Wiktorowicz}, G., {Lasota}, J.-P., {Belczynski}, K., {et~al.} 2021, \apj, 918,
  60, \dodoi{10.3847/1538-4357/ac0cf7}

\bibitem[{{Wiktorowicz} {et~al.}(2019){Wiktorowicz}, {Lasota}, {Middleton}, \&
  {Belczynski}}]{Wiktorowicz2019}
{Wiktorowicz}, G., {Lasota}, J.-P., {Middleton}, M., \& {Belczynski}, K. 2019,
  \apj, 875, 53, \dodoi{10.3847/1538-4357/ab0f27}

\bibitem[{{Wiktorowicz} {et~al.}(2017){Wiktorowicz}, {Sobolewska}, {Lasota}, \&
  {Belczynski}}]{Wiktorowicz2017}
{Wiktorowicz}, G., {Sobolewska}, M., {Lasota}, J.-P., \& {Belczynski}, K. 2017,
  \apj, 846, 17, \dodoi{10.3847/1538-4357/aa821d}

\bibitem[{{Zuo} {et~al.}(2021){Zuo}, {Song}, \& {Xue}}]{Zuo2021}
{Zuo}, Z.-Y., {Song}, H.-T., \& {Xue}, H.-C. 2021, \aap, 649, L2,
  \dodoi{10.1051/0004-6361/202140792}

\end{thebibliography}
\bibliographystyle{aasjournal}

\begin{acknowledgements}
We are grateful to the referee for helpful comments. This work was supported by the National Key Research and Development Program of China (2021YFA0718500) and the Natural Science Foundation of
China under grant Nos. 12041301 and 12121003.
\end{acknowledgements}

\end{document}